\documentclass[a4paper, amsfonts, amssymb, amsmath, reprint, showkeys, nofootinbib, twoside,superscriptaddress,longbibliography]{revtex4-1}
\usepackage[utf8]{inputenc}
\usepackage{graphicx} % For including figures
\usepackage{braket}
\usepackage{siunitx}
\usepackage{amsmath}
\usepackage{amsthm}
\usepackage{color}

\begin{document}
\title{Automated tuning of double quantum dots\\ into specific charge states using neural networks}

\author{R. Durrer}
\affiliation{Department of Physics, ETH Zürich, Otto-Stern-Weg 1, CH-8093 Zürich, Switzerland}
\author{B. Kratochwil}
\affiliation{Department of Physics, ETH Zürich, Otto-Stern-Weg 1, CH-8093 Zürich, Switzerland}
\author{J. V. Koski}
\affiliation{Department of Physics, ETH Zürich, Otto-Stern-Weg 1, CH-8093 Zürich, Switzerland}
\author{A. J. Landig}
\affiliation{Department of Physics, ETH Zürich, Otto-Stern-Weg 1, CH-8093 Zürich, Switzerland}
\author{C. Reichl}
\affiliation{Department of Physics, ETH Zürich, Otto-Stern-Weg 1, CH-8093 Zürich, Switzerland}
\author{W. Wegscheider}
\affiliation{Department of Physics, ETH Zürich, Otto-Stern-Weg 1, CH-8093 Zürich, Switzerland}
\author{T. Ihn}
\affiliation{Department of Physics, ETH Zürich, Otto-Stern-Weg 1, CH-8093 Zürich, Switzerland}
\author{E. Greplova}
\email{geliska@phys.ethz.ch}
\affiliation{Department of Physics, ETH Zürich, Wolfgang-Pauli-Str. 27, CH-8093 Zürich, Switzerland}

\begin{abstract}
While quantum dots are at the forefront of quantum device technology, tuning multi-dot systems requires a lengthy experimental process as multiple parameters need to be accurately controlled. This process becomes increasingly time-consuming and difficult to perform manually as the devices become more complex and the number of tuning parameters grows. In this work, we present a crucial step towards automated tuning of quantum dot qubits. We introduce an algorithm driven by machine learning that uses a small number of coarse-grained measurements as its input and tunes the quantum dot system into a pre-selected charge state. We train and test our algorithm on a GaAs double quantum dot device and we consistently arrive at the desired state or its immediate neighborhood.
\end{abstract}

\maketitle

\section{Introduction}\label{sec:intro}
\paragraph*{} % Intro to quantum devices
Engineered quantum devices are currently at the forefront of scientific research \cite{Lawrie2019, Krantz2019, Prada2019andreev}. The enormous progress in the precision of the engineering and control of these devices at the level of individual particles or excitations allows researchers to build quantum coherent devices such as qubits which are the building blocks of quantum information processing architectures. Such devices have been successfully built utilizing different platforms such as NV-centers \cite{Schirhagl2014,Bradley201910}, superconducting qubits \cite{Kjaergaard2019,Arute2019}, thin nanomaterials \cite{Eich2018spin} and nanowires \cite{Nichele2017,Lutchyn2018}, to name just a few.

\paragraph*{} % intro to QD -> qubit implementation -> tuning as a challenge
Extrapolating the success of semiconductor technology and its scaling properties from the past into the future, qubits based on gate defined semiconductor quantum dots (QDs) are seen as promising candidates for scalable quantum computing. Experiments have successfully demonstrated long coherence times \cite{Kawakami2014}, fast gate operations \cite{Nowack2007,He2019} and long-distance qubit coupling \cite{vanWoerkom2018,Scarlino2019, Landig2019,Borjans2019}. However, to change the number of electrons/holes in the QDs and to tune the coupling to neighbouring QDs and reservoirs, an operator has to change the corresponding voltages applied to the gate electrodes. The relation between applied gate voltages and the physical parameters to be tuned is highly device specific and, thus, requires calibration measurements. Automating the tuning process is therefore one of the key challenges in making semiconductor architectures scalable.

\paragraph*{} % What has been done in auto tuning
Progress in characterizing, controlling and tuning complex quantum systems has been achieved through a variety of algorithmic methods \cite{Kelly2018, Greplova2017,van2018automated,glaser2015training}. Applying machine learning techniques for parameter estimation and tuning of quantum systems has been a promising avenue within this endeavor \cite{Lennon2018,Greplova2017ml,Teske2019,magesan2015machine, Darulova2019,Kalantre2017}. Machine learning methods can be used to automate tasks previously done by humans \cite{Greplova2019fully} and construct high-quality abstract models interpreting complex measurements. They are fast to evaluate even without implementing a device- or system-specific physical model, which can be complex and lengthy to simulate in the case of QD qubits.

\paragraph*{} % What we did
In this work we present a stepping stone towards automated QD tuning. Our algorithm automatically tunes a double quantum dot (DQD) initially in an unknown charge state into a pre-defined charge state. This is relevant as different qubit implementations require a well-defined number of electrons in each QD, such as for the hybrid qubit \cite{Cao2016,Shi2012}, the resonant exchange qubit~\cite{Medford2013,Landig2018}, or the quadrupolar exchange-only qubit~\cite{Russ2018}. We employ convolutional neural networks to recognize transitions between charge states in the charge stability diagram. We test our method by training and testing it experimentally with two DQD devices based on a GaAs heterostructure, and we evaluate its performance.

\section{Experimental Setup}\label{sec:experimental_setup}

\begin{figure}
	\centering
	\includegraphics[width=8.4cm]{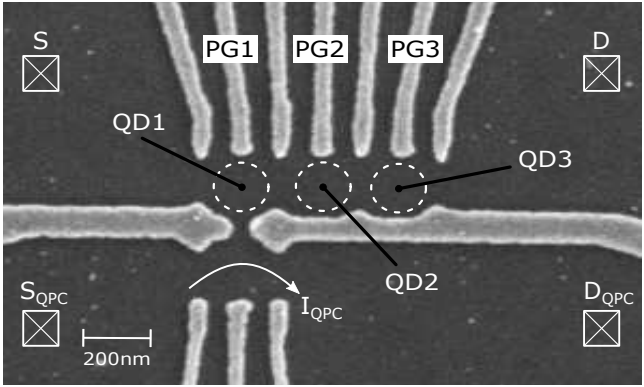}
	\caption{Scanning electron microscope image of the GaAs/AlGaAs sample. The dashed circles indicate the positions of the three QDs that can be formed. The crossed boxes indicate ohmic contacts.}
	\label{fig:sample}
\end{figure}

\paragraph*{} % Device
A scanning electron micrograph of our device, that is capable of forming up to three QDs, is shown in Fig.\ref{fig:sample}. The device has a two-dimensional electron gas \SI{90}{\nano\metre} below the surface, embedded into a GaAs/AlGaAs heterostructure. We confine electrons by applying negative voltages to gold gates on top of the heterostructure. The design permits the formation of up to three QDs (labeled QD1, QD2 and QD3) in a linear array. This allows us to realize two different DQD devices, formed either by QD1 and QD2 (DQD1), or by QD2 and QD3 (DQD2). Whenever we form a DQD the remaining QD is not formed and is therefore part of the reservoir. Three finger gates, visible at the bottom of Fig.\ref{fig:sample}, are used individually to define a quantum point contact (QPC) measuring charge transitions in the nearby QDs. When we measure charge transitions of DQD1 (DQD2) we chose the middle (right) gate as the QPC gate and keep the remaining two gates grounded.

\begin{figure}
	\centering
	\includegraphics[width=8.6cm]{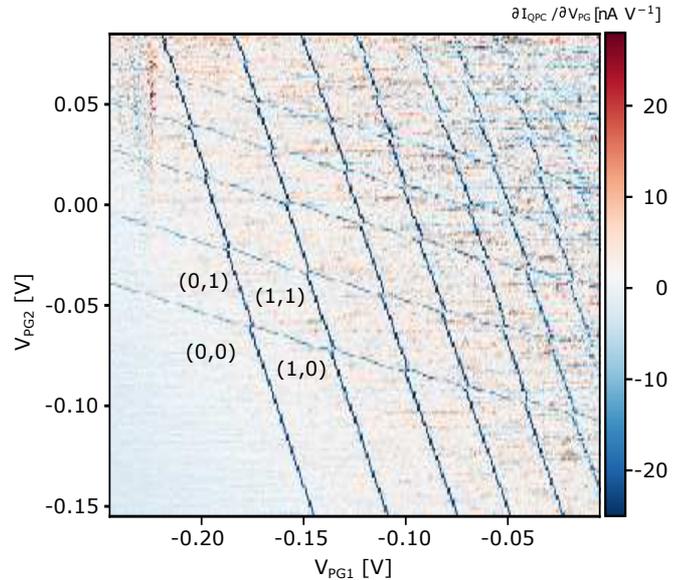}
	\caption{Charge stability diagram of the DQD based on QD1 and QD2 from the sample in Fig. \ref{fig:sample}. The charge occupation is indicated as ($n$,$m$) where $n$ denotes the number of electrons on QD1 and $m$ the number of electrons on QD2.}
	\label{fig:csd}
\end{figure}

\paragraph*{} % Introduce the charge stability diagram
Fig.~\ref{fig:csd} shows a charge stability diagram of DQD1. We plot the change $\partial I_\mathrm{QPC}/\partial V_\mathrm{PG}$ of the current $I_\mathrm{QPC}$ in the quantum point contact as a function of the plunger gate voltages $V_\mathrm{PG1}$ and $V_\mathrm{PG2}$. A change of the electron number in either of the dots causes a sharp peak in $\partial I_\mathrm{QPC}/\partial V_\mathrm{PG}$, seen as tilted lines in Fig.~\ref{fig:csd}. The slopes, intensities and separations of these lines, called charge transition lines, are not precisely reproducible with present day technologies in nominally identical devices and are therefore device-specific. Thus, charge stability diagrams of distinct devices may look different. Furthermore, the properties of the charge stability diagrams may vary for different gate voltages using the same device. This diversity of charge stability diagrams makes manual tuning of such quantum devices a matter of human supervision, where the experience of the operator influences the efficiency of the process.

\paragraph*{} % charge states
We label the charge states as ($n$,$m$) with $n$ and $m$ being the number of electrons in QD1 and QD2, respectively. We assign the charge states to plaquettes of the charge stability diagram by counting the number of charge transition lines starting from the (0,0) state as illustrated in Fig. \ref{fig:csd}. 

\paragraph*{}
A scheme for the efficient measurement of the charge stability diagram driven by machine learning has been introduced in Ref.~\onlinecite{Lennon2018}.
Automated tuning to a specific QD regime (single or DQD) has been recently achieved in Refs.~\onlinecite{Darulova2019} and \onlinecite{Zwolak2019}. Our algorithms complement these achievements and could be combined with some of them.
The starting point for the procedures developed here is any gate voltage setting, where a DQD in the Coulomb-blockade is well-defined, with unknown and arbitrary charge state. The problem we attempt to solve is then stated as: 
Given a DQD in an unknown initial charge configuration ($n_\mathrm{i}$,$m_\mathrm{i}$), find a plunger gate voltage configuration for which the charge configuration is equal to the pre-selected charge state ($n_\mathrm{f}$,$m_\mathrm{f}$) required for operating the qubit.

\section{Algorithm}\label{sec:algorithm}
\paragraph*{} % Intro,  split algorithm into two parts
We split the task of finding the required charge configuration into two steps. Starting from $(n_\mathrm{i},m_\mathrm{i})$ the first step is to find a plunger gate voltage configuration for which the DQD is completely empty [charge state (0,0)]. In the second step, the algorithm loads electrons onto the QDs until the desired charge occupation ($n_\mathrm{f},m_\mathrm{f}$) is reached. For each of the two steps we use a convolutional neural network trained to recognize charge transition lines in the charge stability diagram. A detailed discussion of the machine learning models used is provided in section~\ref{sec:ml}.

\begin{figure*}[]
	\centering
	\includegraphics[width=17.2cm]{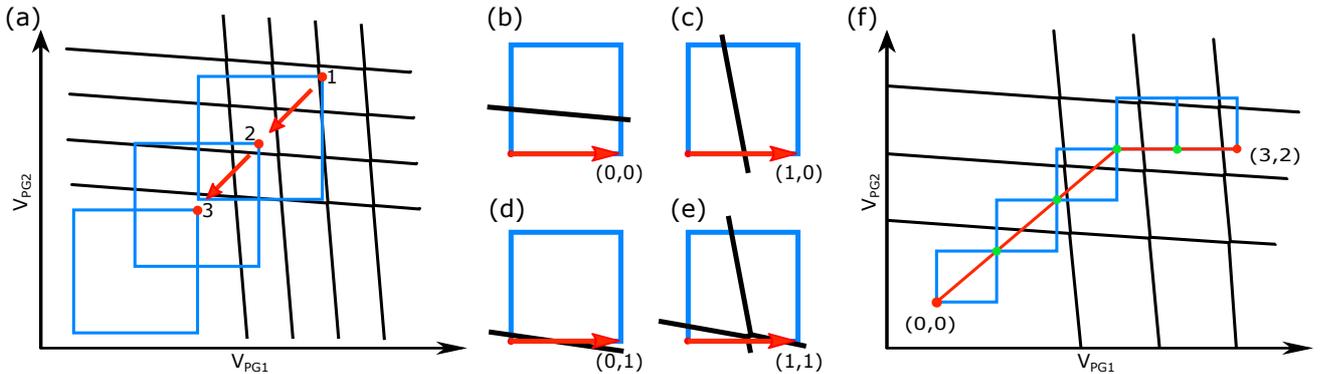}
	\caption{(a) Schematic illustration of the algorithm finding a charge reference point. The black lines indicate charge transition lines, the blue boxes are the patches used for the classification with the neural network. The red dot at the top right corner of a patch corresponds to the point in the charge stability diagram for which the charge occupation is assessed. (b)-(e) Schematics of the patches for the second neural network in blue and charge transition lines in black. The red arrows indicate a path segment. Along each of the indicated path segments a different charge transition occurs indicated as ($n$, $m$) at the end of the arrows. (f) An illustration of a path (red line) connecting the ($0$,$0$) charge regime with the ($3$,$2$) charge regime. The path constitutes of five path segments belonging to the patches in blue. The algorithm chooses the diagonal path when the electron number in both dots needs to be increased. As soon as the final occupation on the second dot is reached only horizontal path segments are used to only load electrons on QD1.}
	\label{fig:algo}
\end{figure*}

\paragraph*{} % Reference charge finding
As illustrated in Fig. \ref{fig:algo} (a), we find the $(0, 0)$ charge
configuration by measuring a small rectangular patches of the
charge stability diagram with low resolution. If the algorithm identifies any charge transition lines it decreases both plunger gate voltages. This process is repeated until no charge transition lines are recognized by the convolutional neural network. Compared to the measurement shown in Fig. \ref{fig:csd} (resolution 1mV, $241 \times 241$ points) the patches are low resolution (resolution 6 - \SI{9}{mV}) and small size ($20 \times 20$ points), and thus fast to measure. However, the size of these patches is chosen at least twice as large as the largest line spacing between two consecutive charge transition lines to ensure that no line is missed. We can determine the patch size experimentally by one dimensional plunger gate sweeps or estimate it from the geometric capacitance between the gate and the dots \cite{vanderWiel2002}. We refer to the plunger gate voltage configuration, for which the algorithm determines the DQD to be empty, as the ``reference point''. This reference point is the starting point for the second step of the tuning algorithm.

\paragraph*{} % Recognizing Charge transitions
In the second step, we tune from the reference point to the desired charge state $(n_\mathrm{f},m_\mathrm{f})$. We do this by again measuring rectangular patches of the charge stability diagram and subsequently classifying them using another convolutional neural network. However, the patches are now $28 \times 28$ measurement points with a finer resolution (of \SI{1}{mV}). Their size is chosen such that at most one charge transition per QD occurs in-between any two corners of the patch. The convolutional network is trained to recognize charge transitions occurring between the lower left corner of the patch and the remaining corners. The four different classification outcomes of this step are illustrated in Figs.~\ref{fig:algo} (b) - (e). We place the lower left corner of the first patch at the $(0,0)$ reference point. We then proceed as depicted in Fig.~\ref{fig:algo} (f). The patches are connected and jointly build a path along which charge transitions are identified. For each corner of the patches, the charge occupation of the DQD is known, and the path consisting of the patches is extended until the pre-selected charge state is found.

\section{Machine Learning Models}\label{sec:ml}
\paragraph*{} % Intro to the methods used
We use two different neural network-based models for the two steps described above, namely to (a) find the zero charge reference point $(0,0)$ and (b) determine the path in the parameter space of plunger gate voltages towards the pre-selected charge configuration $(n_\mathrm{f},m_\mathrm{f})$. Both models are trained to recognize the charge transition lines in the measured patches. The plunger gate voltages are then adjusted based on the presence and direction of the charge transition lines.

\paragraph*{} % Reference point finder
The convolutional neural network for the first step of the algorithm, responsible for finding the $(0,0)$ state, is trained by supervised learning using coarsely measured patches of $20\times 20$ data points in the charge stability diagram as an input. The output of the network is binary: either both QDs are empty or at least one of them is occupied. In other words, since the patch of the charge stability diagram extends beyond the maximal spacing of two charge transition lines, the network effectively detects if there is a charge transition line for any of the QDs within the patch. The model we use for the first part of the algorithm consists of two convolutional and one dense layer. For the full technical description see Appendix~\ref{app:ml}.

\begin{figure}
	\centering
	\includegraphics[width=8.6cm]{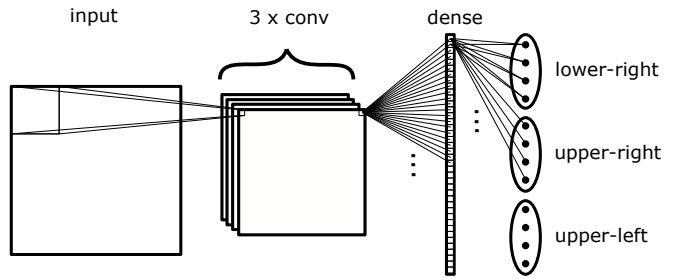}
	\caption{Architecture of the convolutional neural network that recognizes charge transitions between the corners of a given input frame. The neural network consists of three convolutional layers, one dense layer and three outputs with four classification outcomes each.}
	\label{fig:nn_arch}
\end{figure}

\paragraph*{} % charge transition identifier
The neural network for the second part of the tuning-algorithm is trained as a classifier that recognizes individual charge transitions. The output of the network is then used to reach an arbitrary final charge state $(n_\mathrm{f}, m_\mathrm{f})$ from the state $(0,0)$. The input of this more complex classifier are finer $28\times 28$ patches of the charge stability diagram. The output contains $12$ elements that precisely specify which charge transition has been observed in a given patch. More specifically, the presence of one of the possible charge transition scenarios shown in Fig.~\ref{fig:algo} (b) - (e) is determined for the lower right, upper right and upper left corner of the patch (see Fig.~\ref{fig:nn_arch}). Apart from this specific structure of the output design, which we chose to fit to the physics of interest, the body of the network has a standard construction with three convolutional layers and one dense layer. For the technical details of the architecture see Appendix~\ref{app:ml}.

\paragraph*{} % Training
We trained both neural networks on experimentally measured data. In particular, we measured 128 (470) complete charge stability diagrams of DQD1 in finer resolution (coarse resolution) and the operator marked each charge transition line for all data sets. We varied the gate voltage configuration for each of the measured charge stability diagrams and fixed the (compensated) plunger gate range. From each charge stability diagram we cut out random patches, and we use a script to automatically translate the marked transition lines into labels for each patch. For the first part of the algorithm, the labels are single binary variables, while for the second part of the algorithm each patch has three labels, i.e. one for each corner, see Fig.~\ref{fig:nn_arch}. To create a richer training set we augment each charge stability diagram by a factor of $18$. The augmentation is achieved by various combinations of rotations and axis scaling and allows us to include a larger variety of physical plausible measurement scenarios into our training set. For details see Appendix~\ref{app:meas}. The resulting training sets contain the order of $10^5$ patches and both networks are optimized to minimize the error between the labels assigned by the human operator and those assigned by the network. All training data was obtained from DQD1.

\begin{figure}
	\centering
	\includegraphics[width=8.6cm]{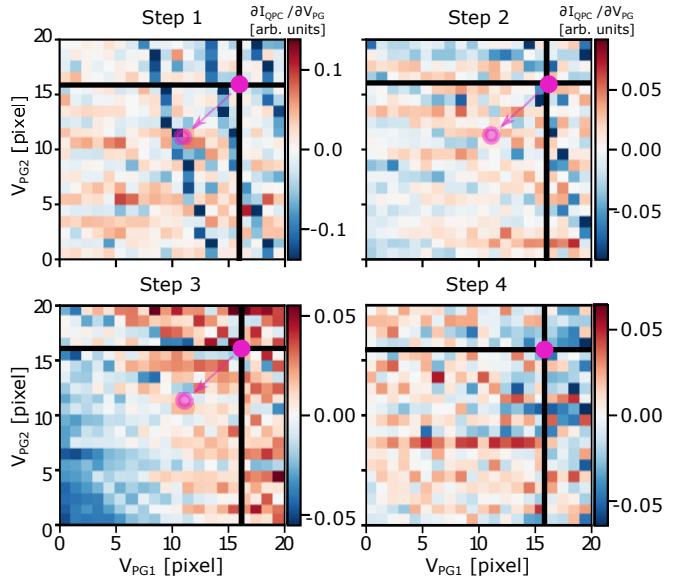}
	\caption{A successful run of the reference finding algorithm. The solid purple dot indicates the voltage configuration for which the charge occupation of the DQD is assessed. The faint purple dot indicates the voltage configuration for which the charge occupation is assessed in the next step. In Step 1 to Step 3, the DQD is occupied, whereas in Step 4 the DQD is empty and the algorithm terminates.}
	\label{fig:ref_run}
\end{figure}

\begin{figure}[b]
	\centering
	\includegraphics[width=8.6cm]{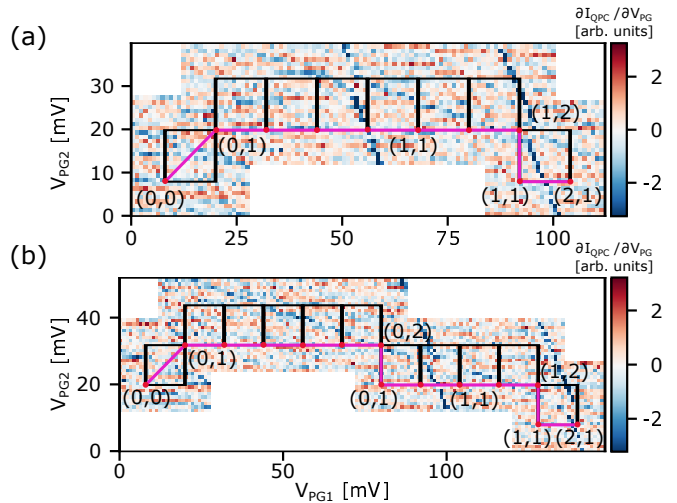}
	\caption{Two successful runs of the auto-tuner for which the plunger gate voltages for the desired charge occupation is found. The boxes in black denote the patches that linked together, form a path shown in red connecting the ($0$,$0$) charge occupation with the desired charge occupation.}
	\label{fig:tran_ran}
\end{figure}

\section{Results} \label{sec:results}
\paragraph*{} % Intro - evaluation on 160 runs
The neural network developed to find the reference point $(0,0)$ is trained on $650'000$ labeled patches. After training the neural network, the model has an accuracy of $98.9\%$ on the evaluation data set. The accuracy score corresponds to the probability that each single patch is classified correctly to represent an empty or an occupied QD. Once the network is able to reliably classify independent patches, we can test the reference point finding part of the algorithm by starting from an arbitrary charge configuration $(n_\mathrm{i}, m_\mathrm{i})$ in the DQD and check if the algorithm reaches the correct reference point $(0, 0)$. The algorithm succeeded in 144 out of 160 runs, resulting in a success rate of $90\%$.
The neural network used for the second part of the algorithm is trained on $530'000$ labeled patches. The accuracy of this neural network, i.e., the probability that the correct charge transition is recognized for any of the corners individually, is found to be $96\%$ on the evaluation data set.

\paragraph*{} % complete runs
We test and evaluate the performance of the complete algorithm by first selecting a desired final state $(n_\mathrm{f},m_\mathrm{f})$. The DQD is then initialized in an arbitrary initial charge state $(n_\mathrm{i}, m_\mathrm{i})$ and we run the first part of the algorithm to reach the reference point $(0,0)$. Once the reference point is reached, the second part of the algorithm identifies the relevant charge transition lines and tunes the DQD to the selected final state.

\paragraph*{} % walk through reference finding
Fig. \ref{fig:ref_run} shows a run of the  first part of the algorithm that identifies the charge reference point. To make sure not to miss any charge transition lines at the edge of the patch, the evaluation of the dot occupation number is done at the crossing point of the two black lines indicated by a purple dot. If the dot is not yet empty a new patch is measured a the position marked by the arrow. In Step 1 to Step 3 there are charge transition lines within the patch and the algorithm correctly classifies them as occupied. In Step 4, no charge transition lines are present and the patch is correctly classified as empty.

\paragraph*{} % walk through electron loading
In Fig.~\ref{fig:tran_ran}, we show two paths connecting a reference point to a point in the charge stability diagram with the correct charge state configuration. Both runs are successful and the correct charge regime $(2,1)$ is found. In black, the charge occupation is indicated after each detected charge transition. Both tuning runs were measured on DQD2 (the system that was not used for training).

\begin{figure}
	\centering
	\includegraphics[width=8.6cm]{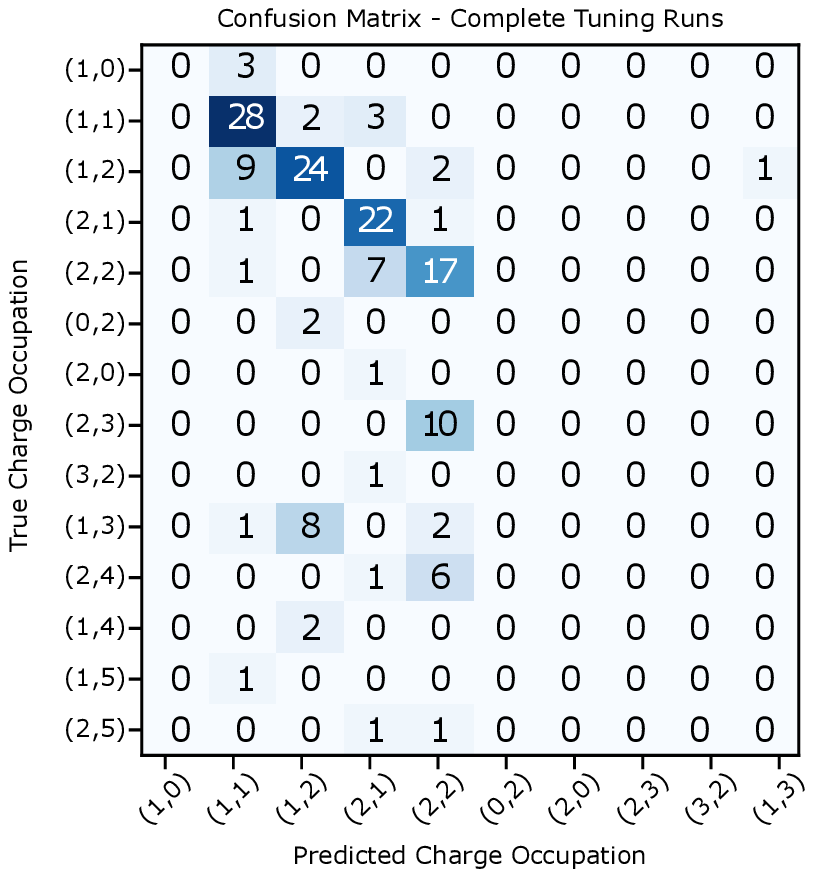}
	\caption{Confusion matrix indicating the outcomes of the test runs of the auto tuner. The rows indicate the actual charge occupation after terminating the algorithm and the column indicates the auto tuners prediction. Two runs are omitted, since their charge occupation could not be determined.}
	\label{fig:conf_matrix}
\end{figure}

\paragraph*{} % Complete tuning runs
The auto-tuner algorithm successfully reached the desired charge configuration in $91$ out of $160$ conducted test runs, which corresponds to a success rate of $57\%$. We tested the tuning on both DQD1 and DQD2 and found the score to be independent of the DQD used. This indicates good generalization of the machine learning models across the device. Allowing for a deviation of one electron, the success rate increases to $89\%$ indicating that the auto-tuner is able to get close to the desired charge regime consistently. We show the confusion matrix of the evaluation of the model in Fig. \ref{fig:conf_matrix}. The entries on the diagonal in the confusion matrix correspond to successful runs, whereas off-diagonal entries represent unsuccessful runs. In particular, one can use off-diagonal entries to identify how exactly each of the erroneous outcomes were misclassified. The off-diagonal entries indicate that most errors occur due to the presence of charge transitions that the algorithm did not recognize in the data. The reverse effect, i.e. identifying a transition that is not physically present in the system, almost never occurs. Thus, we conclude that a weak signal-to-noise ratio in the measurements is the main reason for false classifications.
 The accuracies, $99\%$ and $96\%$ respectively, in identifying the individual transitions are high in the context of neural network classifiers \cite{xiao2017fashion,deng2012mnist}. The total success rate of the auto-tuner is lower than the individual success rate of the neural networks due to cumulative error from the repeated application of the neural networks.

\section{Conclusion and Outlook}\label{sec:conclusion}
\paragraph*{} % 
In this work we implemented and experimentally tested a machine learning assisted auto-tuner that sets the plunger gate voltages to reach any charge configuration of a DQD. We verified that the accuracy of the resulting algorithm persists when testing it on a different DQD than for which the machine learning models were trained on. In our experiments we reached a success rate of around $60\%$. The primary error source is identified as a weak signal-to-noise ratio. We believe this could in principle be alleviated by performing more finely grained measurements for the subsets of the charge stability diagram and, thus, training the convolutional neural networks on higher resolution images. Another suitable strategy is to implement more targeted sampling of the charge stability diagram~\cite{Lennon2018}. Currently, we perform homogeneous steps in voltages at all positions of the chosen patch. Measuring more densely around the places of the diagram where the local standard deviation of the data is maximal would lead to better resolution of the transition lines and might increase the accuracy of the algorithm as well. Further improvements might arise from including simulated data into the training set \cite{Zwolak2019}.

%not sure if this needs to be in the paper, just writing it down for clarity
Generally, neural networks are very good at performing fast and efficient classification, but it is extremely difficult to arrive at $100\%$ accuracy on any training set of practical significance. Thus, the accuracy of individual tuning steps will always have a finite success rate when implemented with a neural network architecture and these finite rates will accumulate if many steps are needed to reach the final state of tuning. These types of auto-tuners may therefore still require some form of human input. Alternatively, additional algorithms that consider the full problem instance, rather than only individual steps, could serve as a better approach.

\section*{Acknowledgements}
We gratefully acknowledge financial support from the Swiss National Science Foundation, the NCCR QSIT. This work has received funding from the European Research Council under grant agreement no. 771503.
We thank Sebastian D. Huber and Klaus Ensslin for fruitful discussions.

\appendix

\begin{figure}
	\centering
	\includegraphics[width=8.6cm]{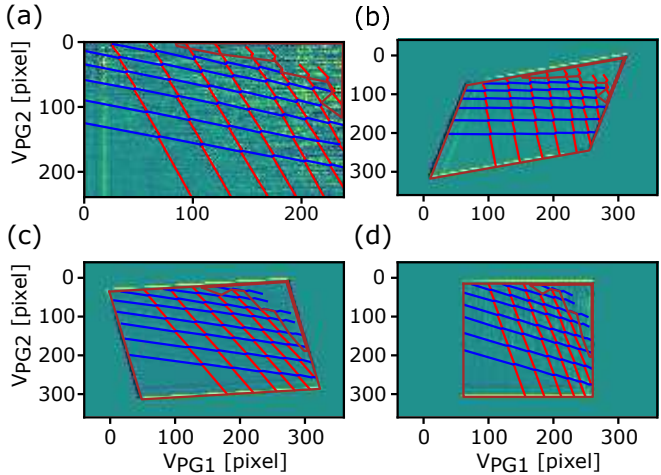}
\caption{(a) Charge stability diagram with indicated charge transitions. Blue lines correspond to charge transition lines of QD2 and red lines to charge transitions of QD1. In the area enclosed by the brown line no charge transition lines could be recognized. (b) - (d) Three different augmentations with randomly chosen transformations of (a).}
	\label{fig:data_aug}
\end{figure}

\begin{table}[t]
	\centering
	%\caption*{Reference Identifier Parameters}
	\begin{tabular}{|c|c|}
		\hline\rule[2.4ex]{0pt}{0pt}
		\textbf{Parameter} & \textbf{Value} \\ \hline\rule[2.4ex]{-4.5pt}{0pt}
		Dropout 1 & 0.05 \\
		Conv. Layer 1 & $(4,4)\times 48$ \\ 
		Activation 1 & ReLu \\
		Dropout 2 & 0.05 \\
		Conv. Layer 2 & $(3,3)\times 12$ \\ 
		Activation 2 & ReLu \\
		Dropout 3 & 0.4 \\
		Dense Layer & 50 \\
		Activation & sigmoid \\
		Outputs & 2 \\
		Activation & softmax \\
		Loss & categorical crossentropy\\	\hline\rule[2.4ex]{0pt}{0pt}
		Optimizer & adam \\
		Batch size & 128 \\
		Learning rate & 0.001\\
		Epochs & 8 \\
		Class weights & True\\	\hline
	\end{tabular}
	\vspace{8pt}
	\caption{Relevant parameters used to create the neural network used to determine charge transition lines in the coarse patches are shown.}
	\label{tab:ref_recogn}
\end{table}

\begin{table}
	\centering
	\begin{tabular}{|c|c|}
		\hline\rule[2.4ex]{0pt}{0pt}
		\textbf{Parameter} & \textbf{Value} \\ \hline\rule[2.4ex]{-4.5pt}{0pt}
		Dropout 1 & 0.2 \\
		Conv. Layer 1 & $(6,6)\times 72$ \\ 
		Activation 1 & ReLu \\
		Dropout 2 & 0.1 \\
		Conv. Layer 2 & $(3,3)\times 24$ \\ 
		Activation 2 & ReLu \\
		Dropout 3 & 0.2 \\
		Conv. Layer 3 & $(2,3)\times 12$ \\ 
		Activation 3 & ReLu \\
		Dropout 4 & 0 \\
		Dense Layer & 50 \\
		Activation & sigmoid \\
		Outputs & $3 \times 4$ \\
		Activation & softmax \\
		Loss & categorical crossentropy\\	\hline\rule[2.4ex]{0pt}{0pt}
		Optimizer & adam \\
		Batch size & 128 \\
		Learning rate & 0.001\\
		Epochs & 8 \\
		Class weights & False\\	\hline
	\end{tabular}
	\vspace{8pt}
	\caption{Relevant parameters used to create the neural network that recognizes individual charge transition between the corners of the high resolution patches are shown.}
	\label{tab:trans_recogn}
\end{table}

% classification results on evaluation data-set
\begin{table}
	\centering
	\begin{tabular}{|c|c|c|c|c|}
		\hline\rule[2.4ex]{0pt}{0pt}
		\textbf{Class} & \textbf{precision} & \textbf{recall} & \textbf{f1} & \textbf{support} \\ \hline\rule[2.4ex]{-4.5pt}{0pt}
		occupied & 0.996 & 0.986 & 0.991 & 2063 \\
		empty & 0.975 & 0.993 & 0.984 & 1089\\ \hline \rule[2.4ex]{0pt}{0pt}
		Micro Avg. & 0.989 & 0.989 & 0.989 & 3152\\	\hline
	\end{tabular}
	\caption{Classification scores for the neural network used to determine charge transition lines in the coarse patches. The numbers are obtained from the evaluation data-set containing 3152 data-points. The micro average corresponds to the overall accuracy, i.e. the fraction of data points that are correctly classified.}
	\label{tab:ml_scores_1}
\end{table}

\section{Measurements and Data Processing}\label{sec:measurements}
\label{app:meas}
\paragraph*{} % Measurement devices
All measurements were recorded in a commercially available dilution refrigerator with a base temperature of 12mK. The electronic temperature is $132\text{mK} \pm 9 \text{mK}$.

\paragraph*{} % Data Augmentation
In Fig. \ref{fig:data_aug} we show a charge stability diagram (a) and three of its random augmentations (b) - (d). The charge stability diagrams have a resolution of 1mV and size $241\text{mV} \times 241\text{mV}$. Subsequently, approximately 300 fine patches are drawn from each of the augmented charge stability diagrams (including their original).

The individual augmentation transformations are: (i) rotation of the whole charge stability diagram, (ii) rotation of 45 degrees followed by scaling the diagram in the PG1 and/or PG2 direction followed by a rotation of $-45$ degrees, (iii) scaling the PG1 axis, (iv) scaling the PG2 axis and (v) flip of the PG1 and PG2 axes.
Flipping the axes simulates measurements for which the plunger gate voltages are swept in reverse order. For all charge stability diagrams, the final augmentation sequence consists of randomly picking one of all transformations with random transformation parameters as e.g. rotation angle or scaling factor.

\paragraph*{}% frame drawing
After augmentations of all charge stability diagrams are performed, the patches are drawn. In order to distribute the patches equally over the charge stability diagram, but still preserve randomness, we decided to draw the patches of size $28 \times 28$ px, where the grid points are randomly shifted by up to five pixels in each direction.

\paragraph*{}% Data processing steps
Each of the patches is processed separately. In the first processing step, the data is rescaled such that the variance equals to one. Secondly, we construct the charge stability diagrams by evaluating the derivative of the measured current $I_{QPC}$ according to the formula
\begin{equation}
\frac{\partial I_{QPC}}{\partial V_{PG}}= \frac{1}{2}  \left(\frac{\partial I_{QPC}}{\partial V_{PG1}}+  \frac{\partial I_{QPC}}{\partial V_{PG2}}\right).
\end{equation}
Subsequently, we remove far outlying measurement points based on a fixed standard deviation criterion.

\section{Machine Learning Models}
\label{app:ml}
\paragraph*{} % Training details, tables
Both machine learning models are constructed in Python using the high-level Tensorflow API Keras \cite{chollet2015keras}. All hyper-parameters which define the architectures are shown in Tab. \ref{tab:trans_recogn} and in Tab. \ref{tab:ref_recogn}. Tables \ref{tab:ml_scores_1}, \ref{tab:ml_scores_2}, \ref{tab:ml_scores_3} and \ref{tab:ml_scores_4} show the performance of the two neural networks on the evaluation data set. The scores we show are the following: recall which is calculated as the number of elements that were evaluated correctly by the network divided by the number of true positives, precision which is the number of elements that were evaluated as positive divided by total number of samples and f1 which is harmonic mean between the recall and precision~\cite{geron2017hands}. The accuracy score, which is the fraction of correctly classified data-points, is given by the micro average of the above scores.

\begin{table}
	\centering
	\begin{tabular}{|c|c|c|c|c|}
	\hline\rule[2.4ex]{0pt}{0pt}
	\textbf{Class} & \textbf{precision} & \textbf{recall} & \textbf{f1} & \textbf{support} \\ \hline\rule[2.4ex]{-4.5pt}{0pt}
	(0, 0) & 0.97 & 0.97 & 0.98 & 1202 \\
	(0, 1) & 0.96 & 0.95 & 0.95 & 643\\
	(1, 0) & 0.97 & 0.93 & 0.95& 157\\
	(1, 1) & 0.83 & 0.83 & 0.83 & 52\\ \hline \rule[2.4ex]{0pt}{0pt}
	Micro Avg. & 0.96 & 0.96 & 0.96 & 2054\\	\hline
\end{tabular}
	\caption{Classification scores for the classification of the charge transitions occurring between lower left and top left corner of the high resolution patches. The numbers are obtained from the evaluation data set containing 2054 data points. The micro average corresponds to the overall accuracy, i.e. the fraction of data points that are correctly classified.}
	\label{tab:ml_scores_2}
\end{table}

\begin{table}
	\centering
	\begin{tabular}{|c|c|c|c|c|}
		\hline\rule[2.4ex]{0pt}{0pt}
		\textbf{Class} & \textbf{precision} & \textbf{recall} & \textbf{f1} & \textbf{support} \\ \hline\rule[2.4ex]{-4.5pt}{0pt}
		(0, 0) & 0.97 & 0.98 & 0.97 & 632 \\
		(0, 1) & 0.95 & 0.96 & 0.95 & 517\\
		(1, 0) & 0.96 & 0.96 & 0.96 & 500\\
		(1, 1) & 0.95 & 0.93 & 0.94 & 405\\ \hline \rule[2.4ex]{0pt}{0pt}
		Micro Avg. & 0.96 & 0.96 & 0.96 & 2054\\	\hline
	\end{tabular}
	\caption{Classification scores for the  classification of the charge transitions occurring between lower left and top right corner of the high resolution patches. The numbers are obtained from the evaluation data set containing 2054 data-points. The micro average corresponds to the overall accuracy, i.e. the fraction of data points that are correctly classified.}
	\label{tab:ml_scores_3}
\end{table}

\begin{table}
	\centering
	\begin{tabular}{|c|c|c|c|c|}
	\hline \rule[2.4ex]{0pt}{0pt}
	\textbf{Class} & \textbf{precision} & \textbf{recall} & \textbf{f1} & \textbf{support} \\ \hline\rule[2.4ex]{-4.5pt}{0pt}
	(0, 0) & 0.98 & 0.99 & 0.98 & 1193 \\
	(0, 1) & 0.90 & 0.87 & 0.89 & 151\\
	(1, 0) & 0.98 & 0.97 & 0.97 & 643\\
	(1, 1) & 0.84 & 0.78 & 0.81 & 63\\ \hline \rule[2.4ex]{0pt}{0pt}
	Micro Avg. & 0.97 & 0.97 & 0.97 & 2054\\	\hline
\end{tabular}
	\caption{Classification scores for the  classification of the charge transitions occurring between lower left and lower right corner of the high resolution patches. The numbers are obtained from the evaluation data set containing 2054 data-points. The micro average corresponds to the overall accuracy, i.e. the fraction of data points that are correctly classified.}
	\label{tab:ml_scores_4}
\end{table}

\bibliography{references}

\end{document}